\documentclass[12pt]{article}
\usepackage{graphicx}

\newcommand\pubnumber{SNSN-323-63}
\newcommand\pubdate{November 23, 2016}

\newcommand{\bbbar}{\mathrm{b\bar{b}}}

\newcommand{\GeV}{\,\mathrm{GeV}}
\newcommand{\TeV}{\,\mathrm{TeV}}

\newcommand{\ifb}{\,\mathrm{fb^{-1}}}

\newcommand{\ttbar}{\mathrm{t\bar{t}}}
\newcommand{\ttH}{\mathrm{t\bar{t}H}}
\newcommand{\ttW}{\mathrm{t\bar{t}W}}
\newcommand{\ttZ}{\mathrm{t\bar{t}Z}}

\def\napoli{University of Belgrade\\
Institute of Physics Belgrade, Pregrevica 118, Zemun, SERBIA}
\def\support{\footnote{Work supported by the Swiss National Science Foundation, Switzerland and
Ministry of Education, Science and Technological Development, Serbia under the Project 171019.}}

\def\Title#1{\begin{center} {\Large #1 } \end{center}}
\def\Author#1{\begin{center}{ \sc #1} \end{center}}
\def\Address#1{\begin{center}{ \it #1} \end{center}}

\newcommand\pubblock{\rightline{\begin{tabular}{l} \pubnumber\\
         \pubdate  \end{tabular}}}
\newenvironment{Abstract}{\begin{quotation}  }{\end{quotation}}
\newenvironment{Presented}{\begin{quotation} \begin{center} 
             PRESENTED AT\end{center}\bigskip 
      \begin{center}\begin{large}}{\end{large}\end{center} \end{quotation}}
\def\Acknowledgements{\bigskip  \bigskip \begin{center} \begin{large}
             \bf ACKNOWLEDGEMENTS \end{large}\end{center}}

\def\beq{\begin{equation}}
\def\eeq#1{\label{#1}\end{equation}}
\def\eeqn{\end{equation}}

\def\beqa{\begin{eqnarray}}
\def\eeqa#1{\label{#1}\end{eqnarray}}
\def\eeqan{\end{eqnarray}}

\let\bar=\overbar

\def\Dslash{\not{\hbox{\kern-4pt $D$}}}
\def\dslash{\not{\hbox{\kern-2pt $\del$}}}

\def\msb{{\bar{\ssstyle M \kern -1pt S}}}

\usepackage{amsmath}

\begin{document}
\begin{titlepage}
\pubblock

\vfill
\Title{$\ttH$ production at $13\TeV$}
\vfill
\Author{ Predrag Cirkovic\support on behalf of the CMS Collaboration}
\Address{\napoli}
\vfill
\begin{Abstract}
In this paper, the latest results of searches for the standard model Higgs boson produced in association with a top quark-antiquark pair ($\ttH$), where Higgs decays into photons, bottom quark-antiquark pair or leptons via $\mathrm{W^{+}W^{-}}$, $\mathrm{ZZ^{\left(*\right)}}$ and $\mathrm{\tau^{+}\tau^{-}}$ are presented. The analyses have been performed using the $13\,\mathrm{TeV}$ pp collisions data recorded by the CMS experiment in 2015 and part of 2016. The results are presented in the form of the best fit to the signal strength ($\mathrm{\mu=\sigma/\sigma_{SM}}$) measured with respect to the Standard Model prediction and its expected and observed 95\% C.L. upper limits.
\end{Abstract}
\vfill
\begin{Presented}
The $\mathrm{9^{th}}$ International Workshop on Top Quark Physics\\
TOP 2016\\
Olomouc, Czech Republic,  September 19--23, 2016
\end{Presented}
\vfill
\end{titlepage}
\def\thefootnote{\fnsymbol{footnote}}
\setcounter{footnote}{0}

\section{Introduction}

One of the most important recent experimental discoveries in particle physics has been achieved in the search for a Higgs boson, finding such a particle with a mass around $125\GeV$~\cite{RefHiggs}.
Since then, the focus is on the precise measurement of the Higgs properties, such as its coupling to the top quark.
The top quark is the heaviest fermion and both experiments, CMS and ATLAS at the Large Hadron Collider at CERN, are measuring this coupling directly, via the associated production in the $\ttH$ process.
This process provides a direct access to the top Yukawa coupling, that is indirectly available from the dominant Higgs production mechanism at the LHC, gluon-gluon fusion, with the assumption that no additional (BSM) particles contribute to the production loop.
Therefore, this direct measurement of top Yukawa coupling in the associated production can be used to set constraints on the new physics with respect to the ggH process.
Given the large mass of the top quark, this coupling is large and the top quark is expected to be the main responsible for the instability of the Higgs mass against radiative corrections, therefore playing a role in the electroweak symmetry breaking (EWSB).
The result of the combined CMS and ATLAS measurement of the $\ttH$ production with the 7 and 8 TeV Run I LHC data has achieved the observed (and expected) significance of 4.4\,(2.0)$\sigma$.
The search has been performed in the $\ttH$ ($\gamma\gamma$), $\ttH$ ($\bbbar$) and $\ttH$ (multilepton) final states.

\section{Analysis}

In this report, the first studies of $\ttH$ production at 13 TeV from the CMS Collaboration, using the dataset recorded in 2015 by the CMS experiment is presented.
The considered final states correspond to the decays of Higgs boson to pair of photons, pair of bottom quarks, or the final state with leptons from the decays to WW, ZZ and $\tau\tau$.
The final result represents the combination of the results of the three separately performed measurements.
Although having low cross sections, these processes provide clean signatures for studying the associated production of top quark and Higgs.
With the rise of the center of mass energy from $8\TeV$ to $13\TeV$, the cross section of the $\ttH$ process increases by a factor of 4, while the main backgrounds increase by a factor of 3, leading to an enhanced sensitivity of the measurement at the energy of $13\TeV$.
The common dominant background process in all three analyzes is the $\ttbar$ production in association with vector bosons or jets.
After the final selection, if background yields still dominate the signal, in order to achieve better signal over background ratio, the events are further categorized.
This additional categorization has been performed in the analyzes of the $\ttH$ ($\bbbar$) and $\ttH$ (multilepton) final states.

\subsection{$\mathrm{t\bar{t}H\left(H \rightarrow b\bar{b}\right)}$}

\begin{figure}[ht]
\centering
\includegraphics[width=5.5cm,clip,angle=0,origin=c]{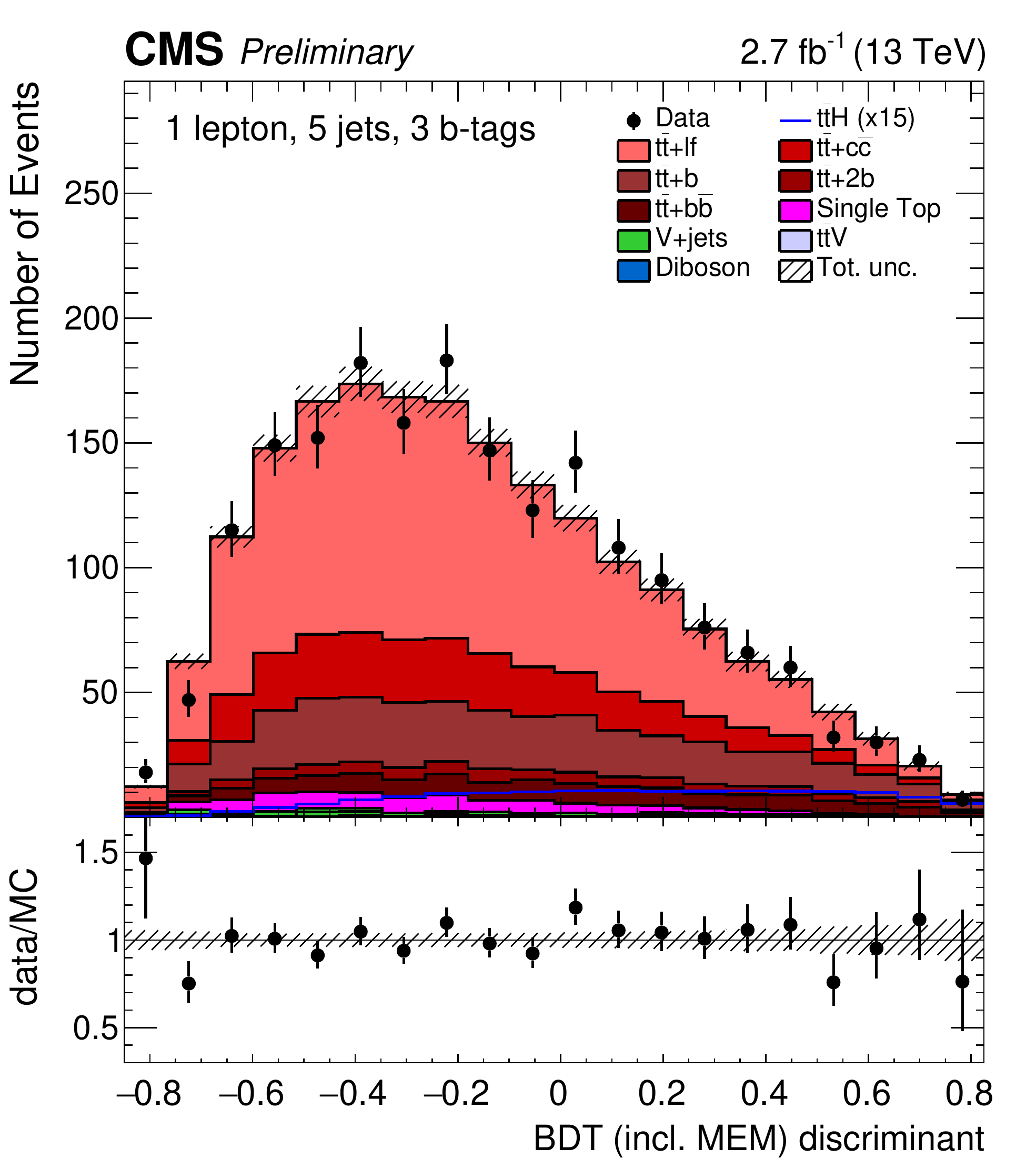}
\includegraphics[width=5.5cm,clip,angle=0,origin=c]{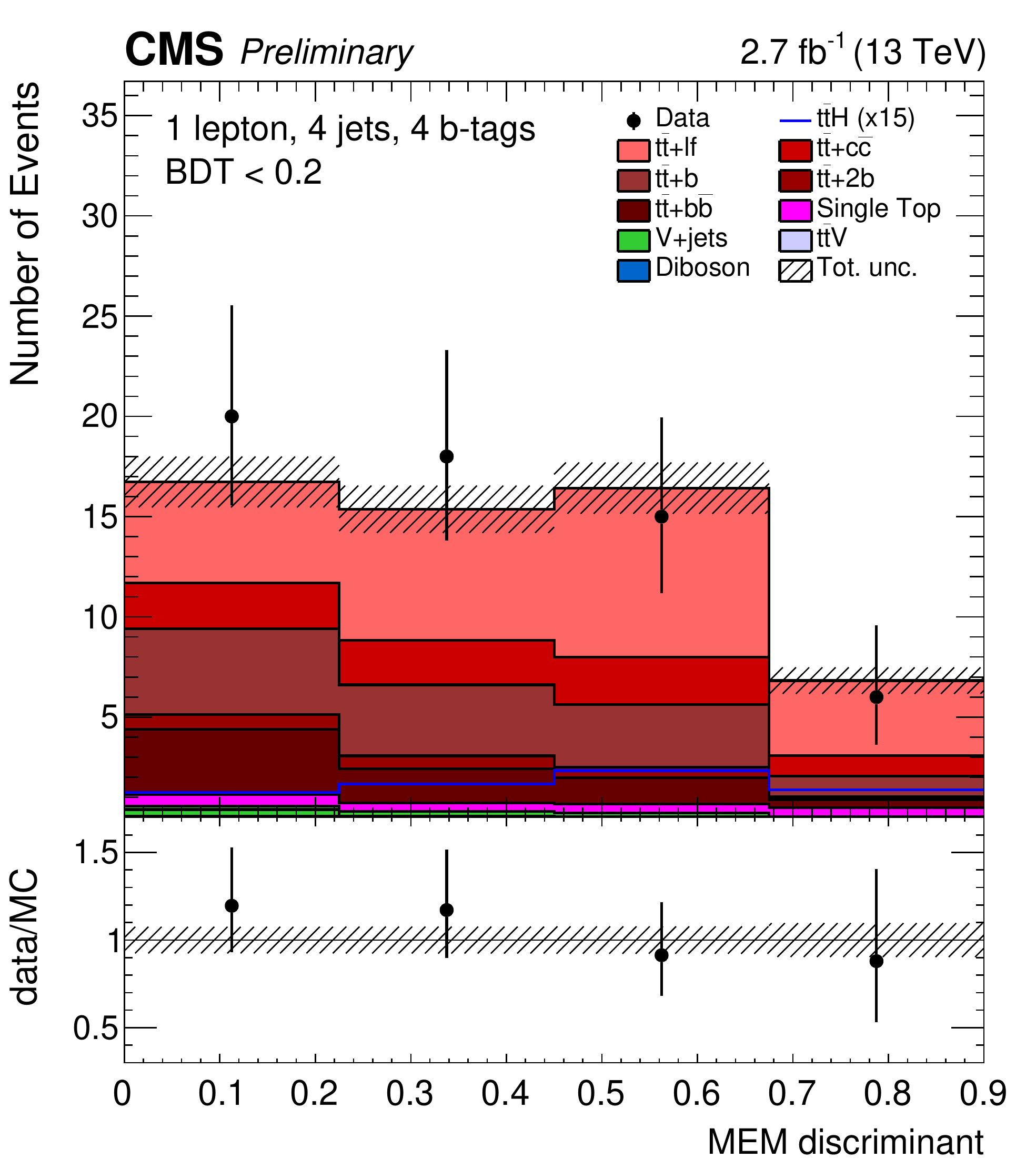}
\caption{
The distribution of the BDT output that includes also the MEM discriminator as one of its input variables, for the 1 leptons, 5 jets of which 3 are b-tagged (left).
The distribution of the MEM discriminator used exclusively in 1 lepton, 4 jets (all b-tagged) category (right).
The background histograms are stacked and normalized to the luminosity of 2.7 $\ifb$~\cite{RefTTHbb}.
}
\label{fig:Hbb}       
\end{figure}

The channel with Higgs decayng to $\bbbar$ pair has the largest branching ratio of 0.56$\pm$0.02 for the Higgs boson mass of 125 GeV~\cite{RefTTHbb}.
The dominant background process here is the $\ttbar$+jets, with the irreducible $\ttbar$+$\bbbar$ as the one that is most difficult to distinguish from signal.
The experimental challenges are also the limited mass resolution for H $\rightarrow$ $\bbbar$ and the presence of jets with similar kinematical properties as the $\ttH$ signal.
The jets associated to the $\ttbar$ pair are split according to their flavour and treated separately in the analysis.
In order to enhance the sensitivity of the analysis, the events are categorized into two main categories defined as ''lepton+jets'', that has exactly one lepton and ''dilepton'', that has exactly two leptons of the opposite sign in the final state.
The ''lepton+jets'' category uses the ability to trigger on the presence of one lepton and therefore to suppress the QCD background, while the ''dilepton'' category provides minimal contribution of the backgrounds other than $\ttbar$ and the reduced jet combinatorics.
As the signal events usually have more (b-tagged) jets than the background, the events are further categorized according to the number of jets and number of b-tagged jets.
The category with high-$p_T$ boosted jets from hadronic decays of top quark or H $\rightarrow$ $\bbbar$ are introduced into $13\TeV$ analysis for the first time.
The ''lepton+jets'' category is broken down into eight subcategories, while the ''dilepton'' category is separated into five subcategories.
In order to achieve the best separation between signal and background, in each of the categories, different discriminators are used: a boosted decision tree (BDT, a machine learning technique), a matrix element method (MEM, a physics-motivated technique), or a combination of the two.
E.g., for the 1 lepton category with 5 jets of which 3 are b-tagged, the final discriminator is obtained using the BDT algorithm taking the MEM weight as one of the inputs.
Its output distribution is shown in Fig.~\ref{fig:Hbb}-left.
The category with 1 lepton and at least 4 jets of which all 4 are b-tagged uses another approach (MEM distribution after cut in BDT discriminant) chosen using the best expected performance in this channel (Fig.~\ref{fig:Hbb}-right).
A simultaneous maximum-likelihood fit is performed to the data in all the categories.
The final results are represented as the signal strength relative to the SM expectation, $\mu = \sigma/\sigma_{SM}$, which is in ''lepton+jets'' category measured to be $-4.7\substack{+3.7 \\ -3.8}$, while in the ''dilepton'' category it is $-0.4 \pm 2.1$.
The combined result yields a signal strength of $-2.0 \pm 1.8$, that is $1.7\sigma$ below the SM expectation.

\subsection{$\mathrm{t\bar{t}H\left(H \rightarrow \gamma\gamma\right)}$}

The smallest branching ratio considered if that of the H $\rightarrow \gamma\gamma$ channel, but it offers a very clean signature~\cite{RefTTHgg1, RefTTHgg2},
since the CMS detector provides an excellent photon identification and energy resolution enabling very precise reconstruction of the diphoton invariant mass peak.
All H $\rightarrow \gamma\gamma$ events are categorized according to the production mode using the the Vector Boson Fusion (VBF) and $\ttH$ taggers.
In terms of the $\ttH$ analysis here are considered the diphoton events with two bottom quarks and two W bosons originating from the decay of $\ttbar$.
The final state may also consider leptons and additional jets.
This analysis used 12.9 $\ifb$ of the CMS data, preselected with diphoton triggers~\cite{RefTTHgg2}.
The dominant background components are $\ttbar$+$\gamma\gamma$ and $\ttbar$+$\gamma$+jet, where the jets are mis-identified as photons.
One of the main tasks of the analysis is to reject the mis-identified jets (fakes).
The events are categorized into the following categories: ''leptonic tag'' and ''hadronic tag''.
''Leptonic tag'' requires at least two jets of which at least one is b-tagged, while ''hadronic tag'' category requires at least five jets with at least one b-tagged jet.
Photon identification is performed using a multivariate analysis based on BDT algorithm.
In each of the two categories, another MVA is used for the selection of the reconstructed diphoton events.
The signal extraction strategy is based on the fit of the combined diphoton mass resonance and the non-resonant monotonically decreasing background.
The signal model fit is performed analytically in each category and for each simulated value of the Higgs mass.
The signal model for the Higgs boson mass of 125 GeV is shown in Fig.~\ref{fig:Hgg}-first row, while the combined signal plus background model is shown in Fig.~\ref{fig:Hgg}-second row, obtained with the events from the ''leptonic tag'' (left) and ''hadronic tag'' (right) category.
The 1 $\sigma$ and 2 $\sigma$ bands (green and yellow, respectively) represent the uncertainty of the fitted background component.
In the ''hadronic tag'' category, a small excess of events is observed close to the Higgs mass, while the ''leptonic tag'' contains fewer events passing the final selection.
The results of the analysis in these two categories for the $m_H = 125.09\GeV$ are signal strengths measured as $2.1\substack{+1.6 \\ -1.2}$ in ''hadronic tag'' and $1.15\substack{+2.0 \\ -1.4}$ in the ''leptonic tag'' category~\cite{RefTTHgg2}.

\begin{figure}[ht]
\centering
\includegraphics[width=5.0cm,clip,angle=0,origin=c]{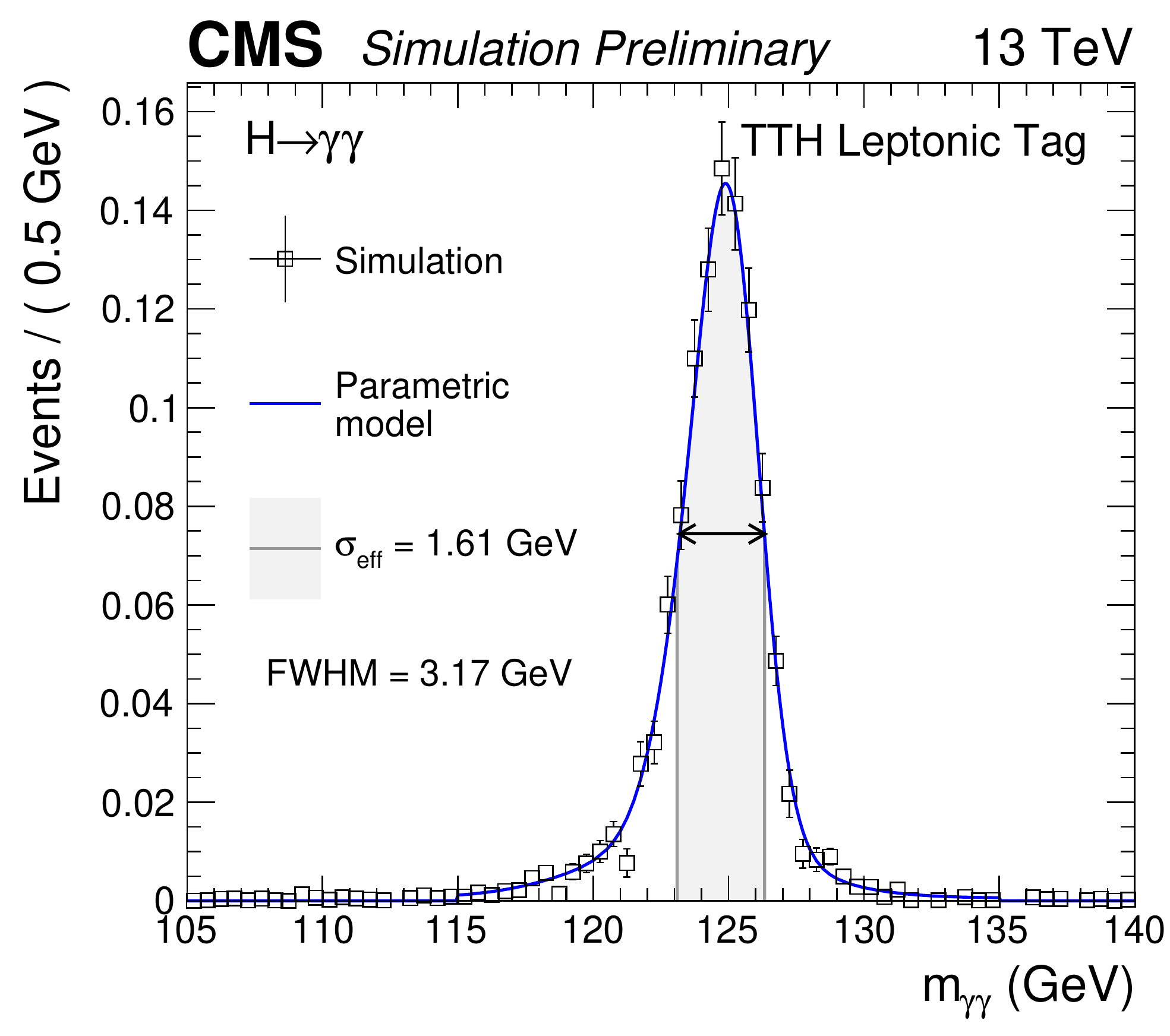}
\includegraphics[width=5.0cm,clip,angle=0,origin=c]{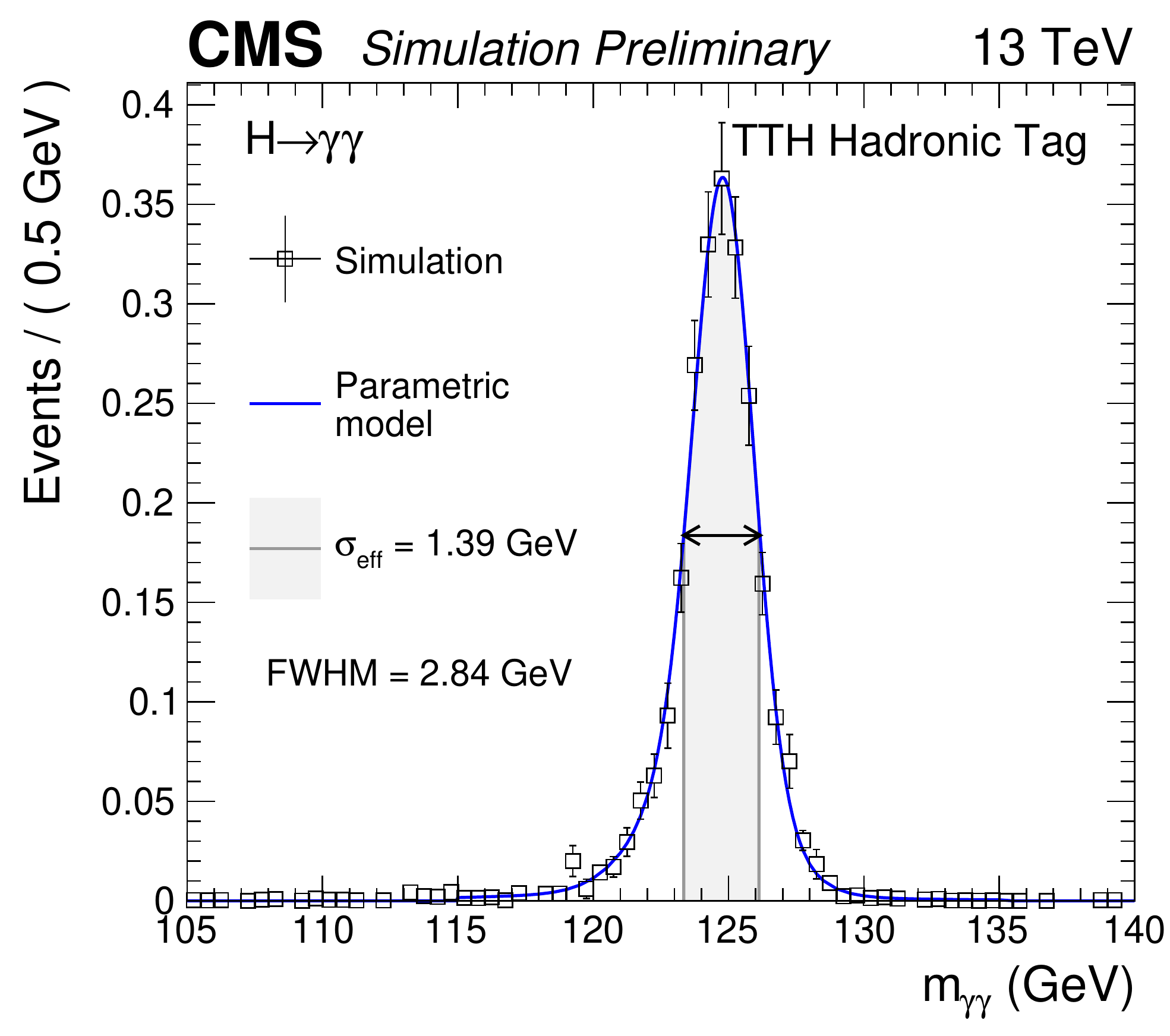}
\includegraphics[width=5.0cm,clip,angle=0,origin=c]{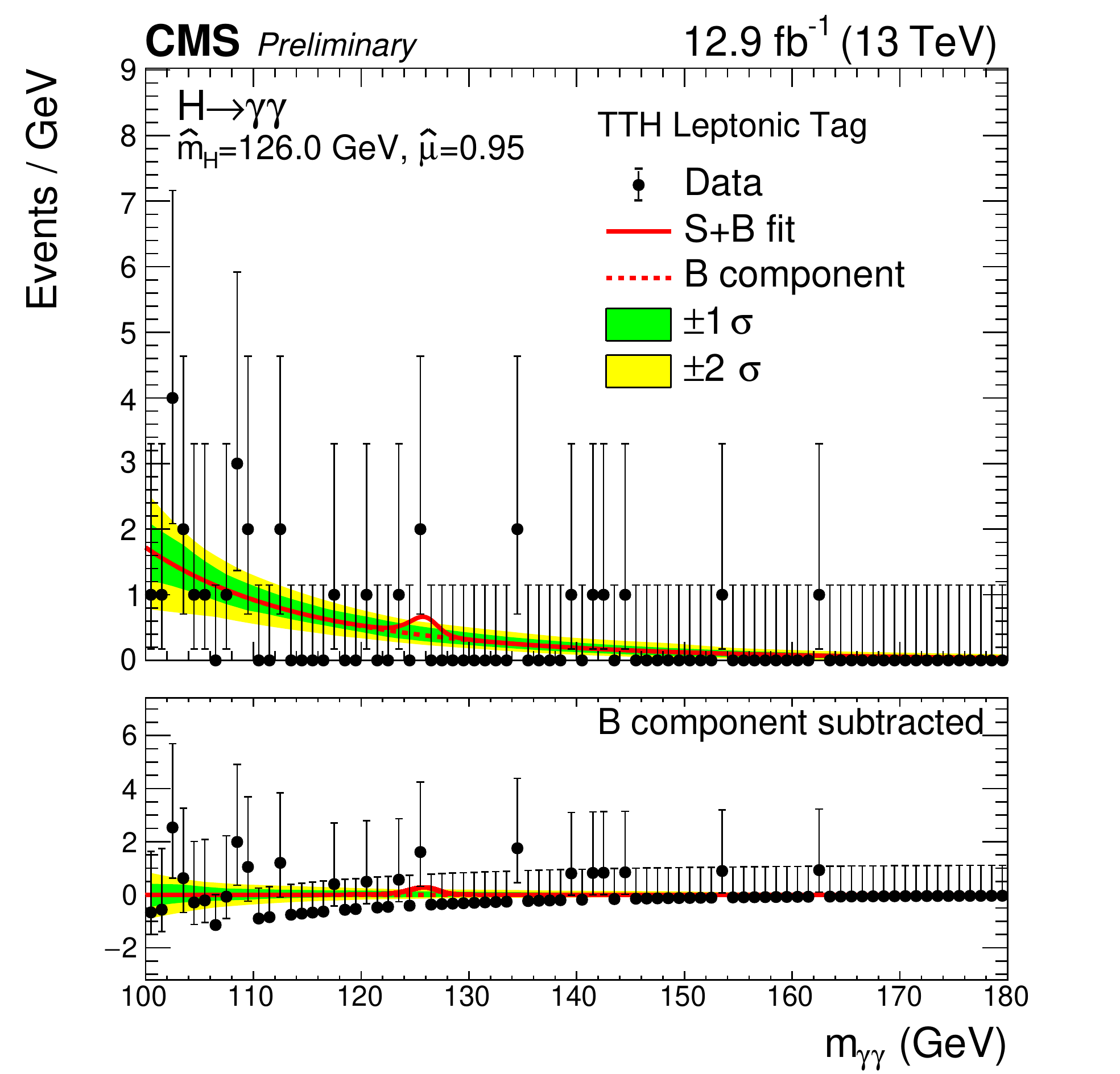}
\includegraphics[width=5.0cm,clip,angle=0,origin=c]{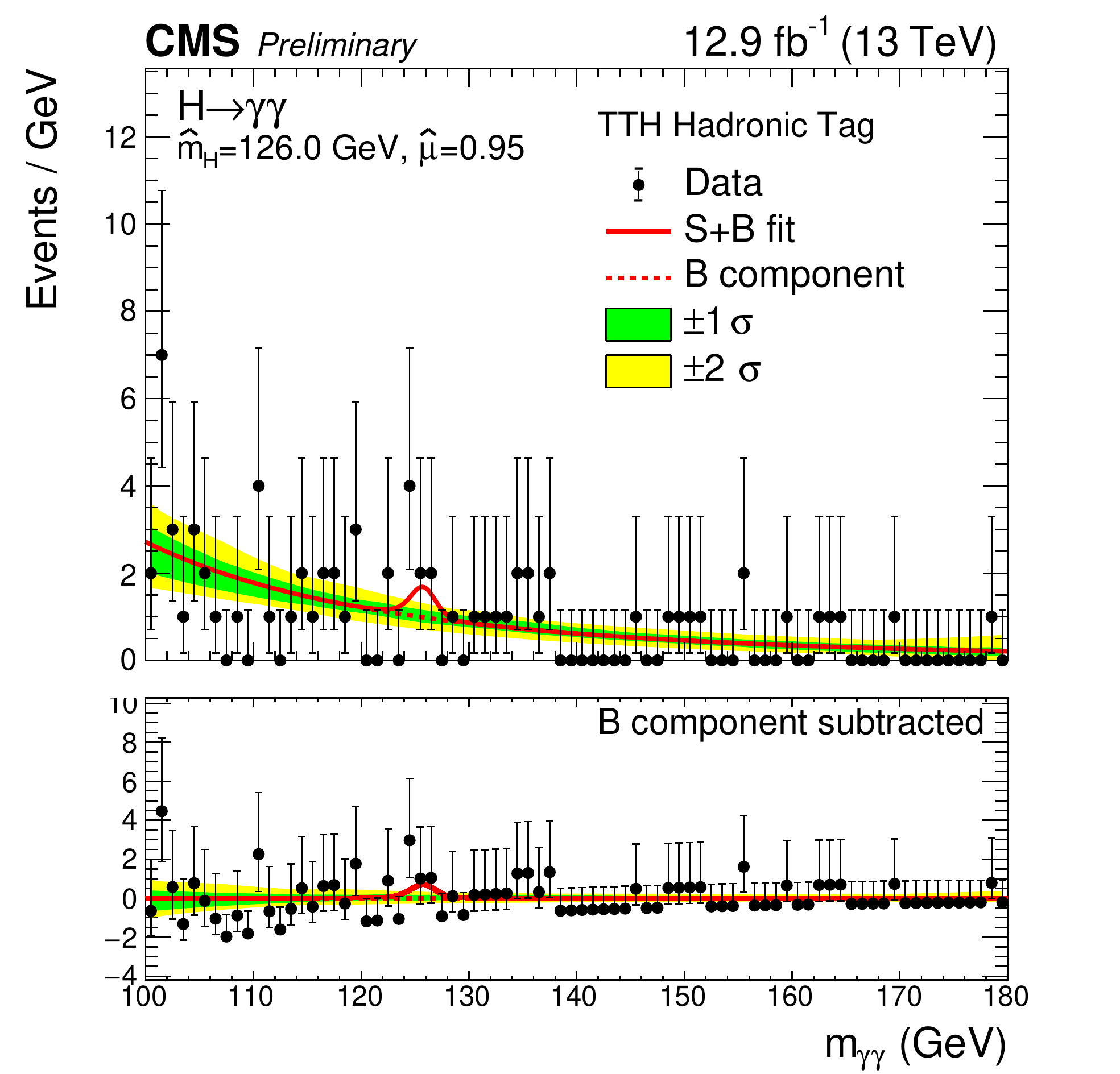}
\caption{
The signal model fit for the ''leptonic tag'' (upper-left) and the ''hadronic tag'' (upper-right) and Higgs boson mass of 125 GeV, and the results of the corresponding signal plus background fit (lower-left and lower-right)~\cite{RefTTHgg2}.
}
\label{fig:Hgg}       
\end{figure}

\subsection{$\mathrm{t\bar{t}H \rightarrow multileptons}$}

Higgs decays to WW, ZZ or $\tau\tau$ where at least one of the vector bosons or $\tau$-leptons decays to leptons are considered in the multilepton analysis~\cite{RefTTHml1, RefTTHml2}.
This channel has a small branching ratio but the presence of leptons from top quark decays provides clean experimental signatures categorized in
two leptons (electrons or muons) with the same sign of the charge, or at least three leptons plus b-tagged jets in the final state.
Two main irreducible background components are $\ttW$ and $\ttZ$ (estimated from the MC) and the reducible component $\ttbar$+jets (estimated using a data-driven fake rate method).
The fake rate is estimated from QCD multijet and Z+jets events.
Control regions are defined to model the non-prompt lepton background by relaxing the lepton identification requirement.
The selected events are accordingly weighted using the function of the mis-identification probability of the leptons.
Contamination of the electrons with mis-identified charge is evaluated from a sample of electrons from Z decays, split into opposite-sign and same-sign pairs.
The probability of the charge mis-reconstruction is measured as a function of $|p_{T}|$ and $|\eta|$ and it ranges from 0.03\% in barrel and 0.4\% in the end-cap region of the CMS detector~\cite{RefTTHml2}.

Single, double and triple lepton (electron or muon) triggers are used for the event preselection.
The two-lepton same-sign category requires at least four additional jets, while the category with at least three leptons requires at least two additional jets.
These two categories are further broken down by the lepton flavour, presence of hadronically decaying $\tau$-leptons and the presence of two b-tags, into fifteen final categories.
In order to maximize the presence of signal with respect to the $\ttbar$ and $\ttW$/Z processes, two separate MVAs are trained.
The two discriminators are used to define bins (in the 2D space) in which a counting experiment is performed.
The distribution of the number of counts in the 2D bins, in the two-lepton same-sign and at least three leptons categories, is shown in the Fig.~\ref{fig:Hml}-left and Fig.~\ref{fig:Hml}-right, and the corresponding best fits of the signal strength are measured to be
$2.7\substack{+1.1 \\ -1.0}$ and $1.3\substack{+1.2 \\ -1.0}$.
The result obtained combining the results from the separate categories is $2.0\substack{+0.8 \\ -0.7}$ (including 2015 CMS data as well)~\cite{RefTTHml2}.
The deviations are not significant enough to show disagreement with the standard model prediction.

\begin{figure}[ht]
\centering
\includegraphics[width=5.0cm,clip,angle=0,origin=c]{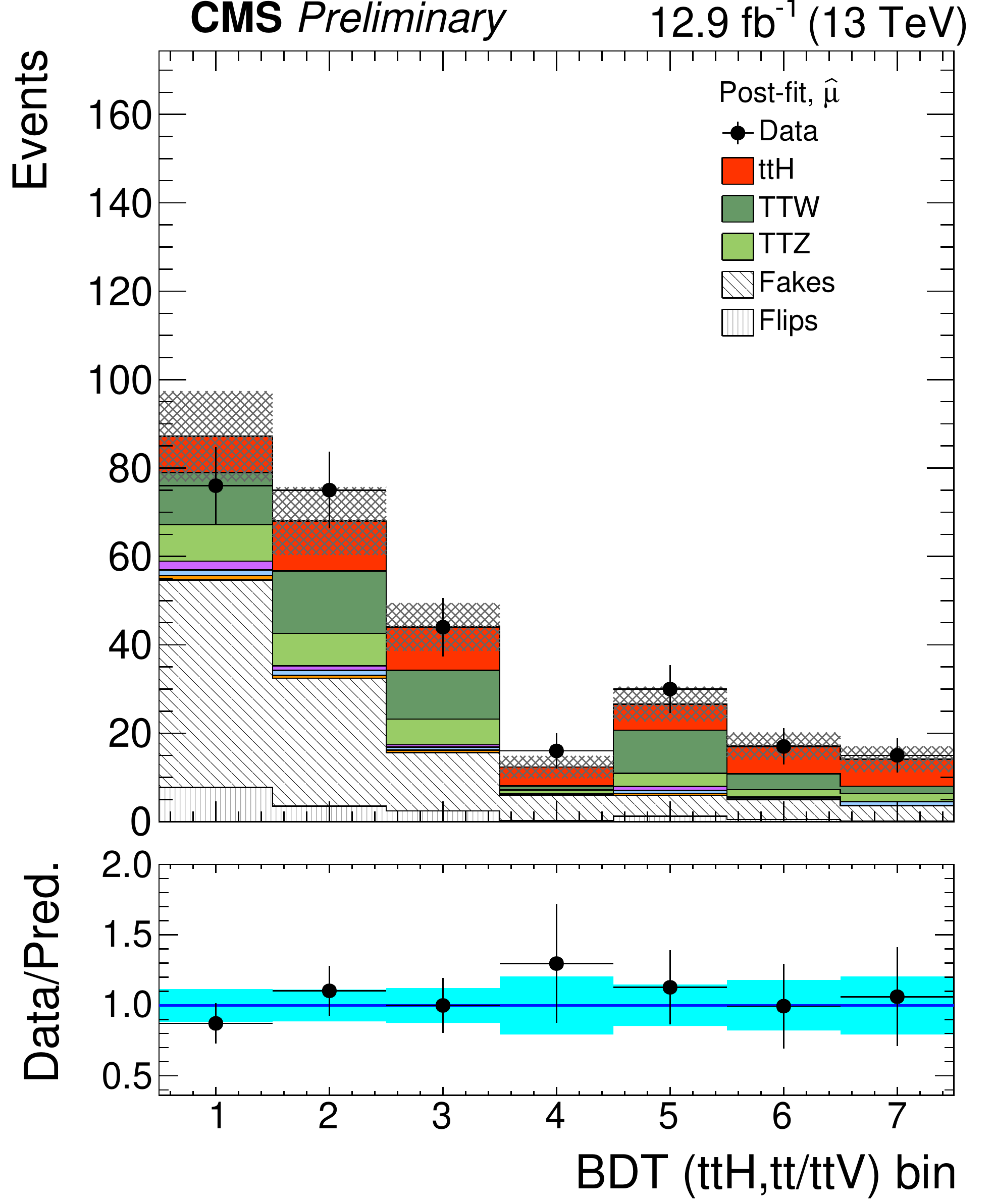}
\includegraphics[width=5.0cm,clip,angle=0,origin=c]{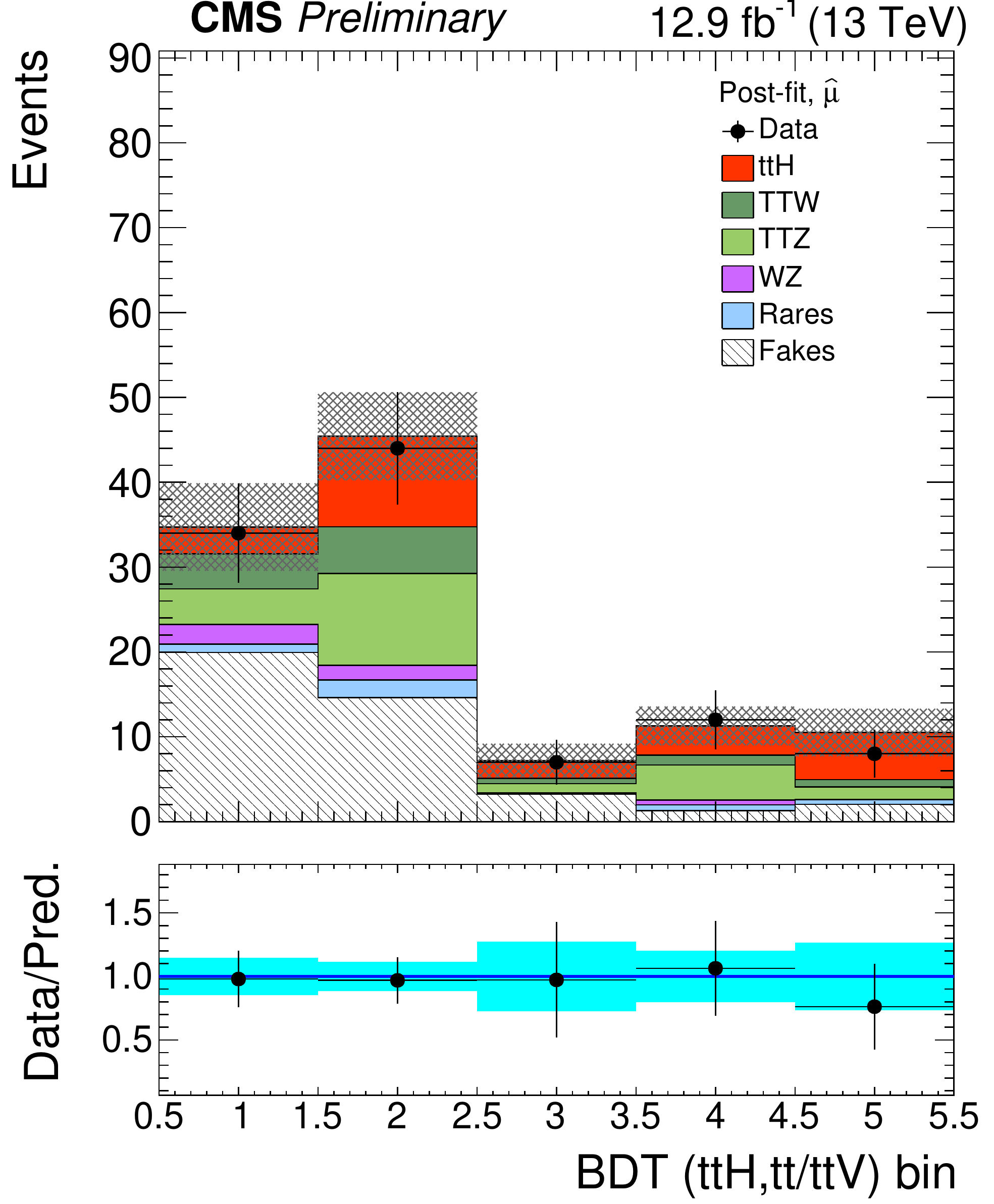}
\caption{
Distribution of the final BDT discriminator, obtained after combining the two BDT trainings against $\ttbar$ and $\ttW$/Z processes, shown for the two-lepton same-sign category (left) and at least three leptons category (right)~\cite{RefTTHml2}.
}
\label{fig:Hml}       
\end{figure}

\subsection{Combined results}

The three channels are combined, for the 2015 data set, assuming the mass of Higgs boson of $125\GeV$ and taking into account the correlation of the common systematic uncertainties~\cite{RefWeb}.
The obtained best fit on the signal strength of $0.15\substack{+0.95 \\ -0.81}$ is in agreement with the SM prediction, which is shown in Fig.~\ref{fig:comb}-left.
The 95\% C.L. limits on the signal strength in each of the analysis channels and their combination, which observed (expected) value is 2.1 (1.9), are shown in Fig.~\ref{fig:comb}-right.

\begin{figure}[ht]
\centering
\includegraphics[width=6.0cm,clip,angle=0,origin=c]{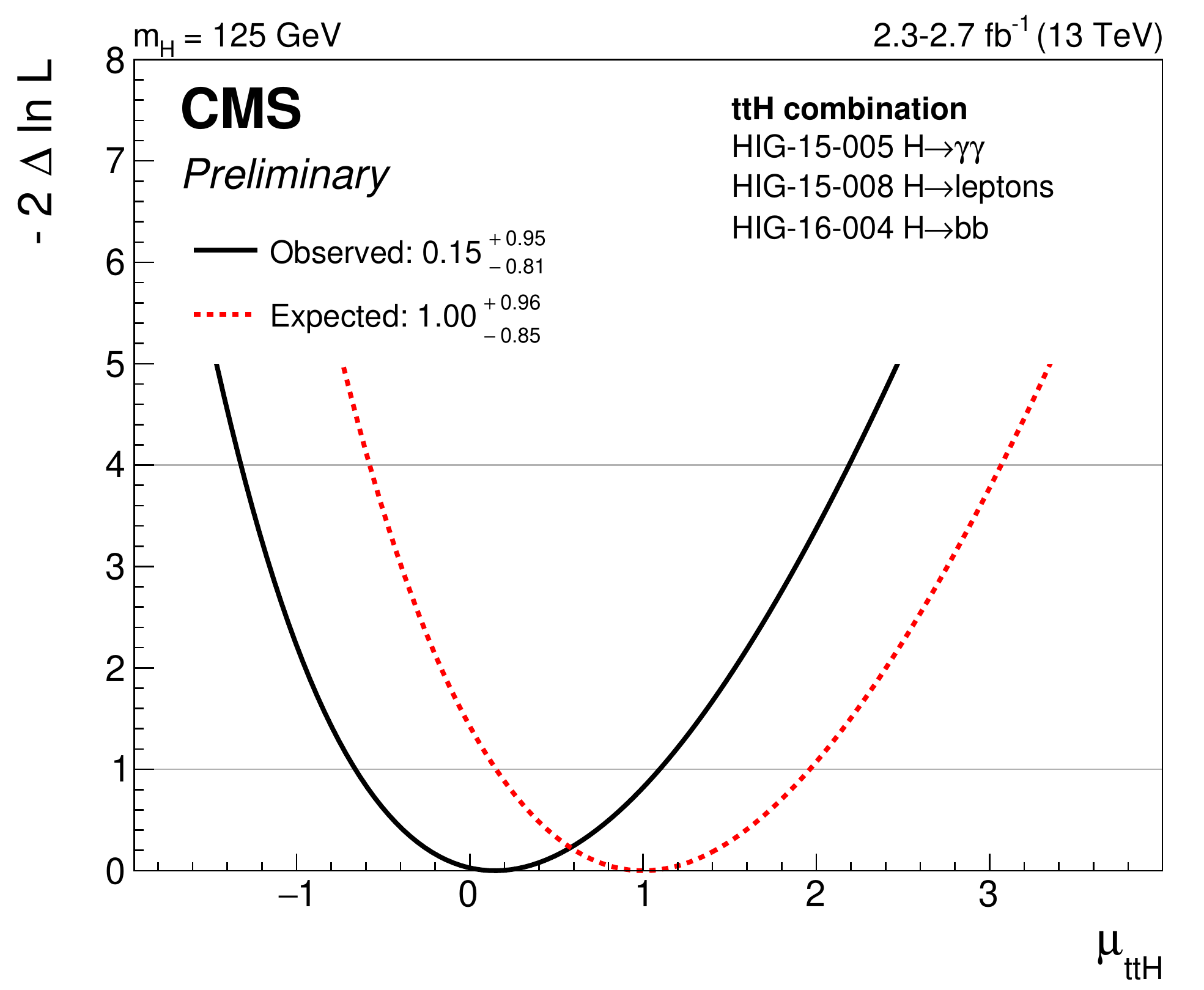}
\includegraphics[width=6.0cm,clip,angle=0,origin=c]{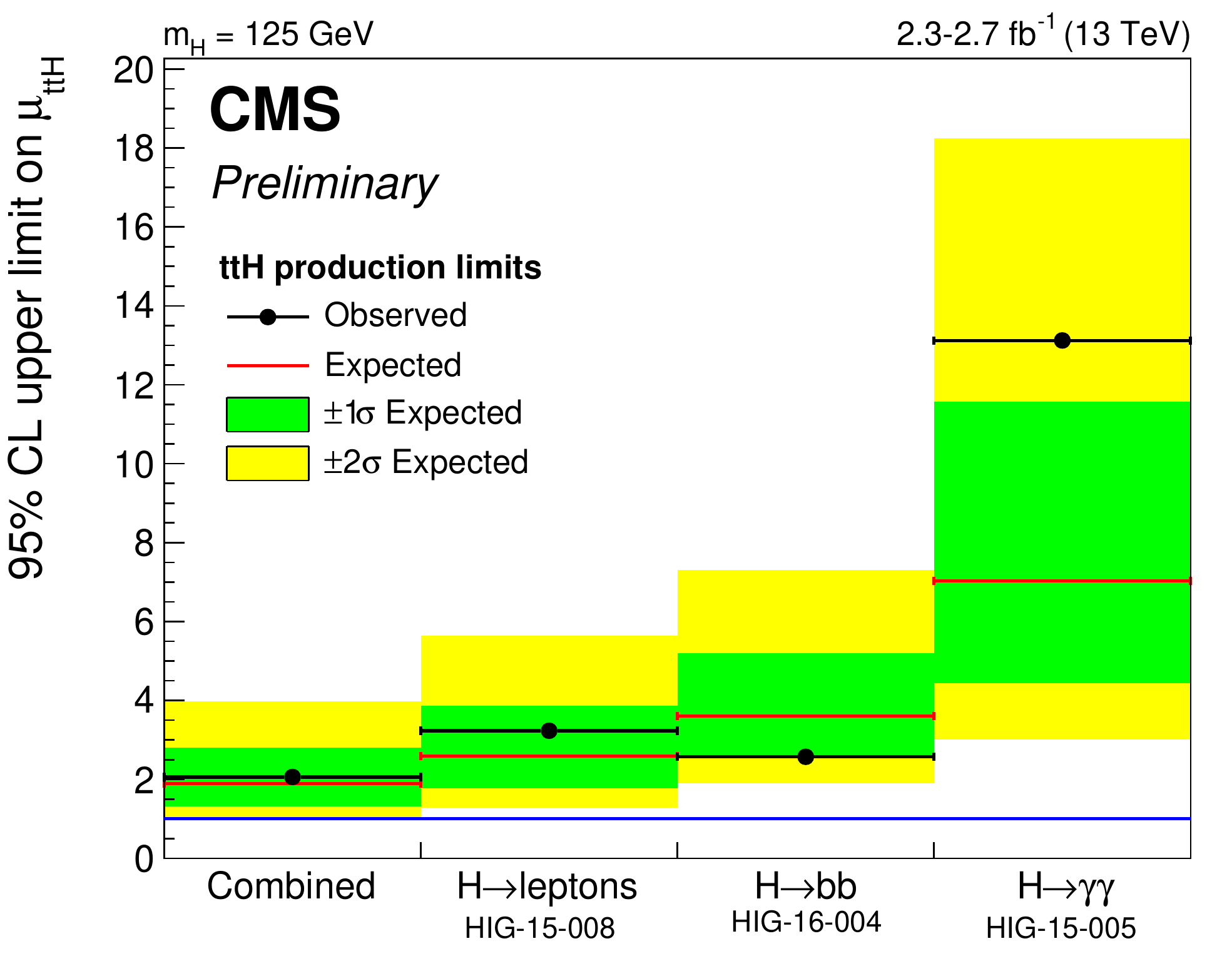}
\caption{
Signal strength measurement of the $\ttH$ production process, $\mu_\ttH$ (left) and the 95\% C.L. upper limits on the rate of $\ttH$ production relative to the standard model expectation ($\mu_\ttH = 1$), combined from the three analysis channels: $\ttH$ ($\gamma\gamma$), $\ttH$ ($\bbbar$) and $\ttH$ (multilepton)~\cite{RefWeb}.
}
\label{fig:comb}       
\end{figure}

\section{Summary}

The first results of the search for the $\ttH$ production at 13 TeV using the 2015 and part of 2016 CMS data have been presented here.
They are obtained from independent channels: $\ttH$ ($\gamma\gamma$), $\ttH$ ($\bbbar$) and $\ttH$ (multilepton), analysed separately and then combined.
Combined best fit to the signal strength of $0.15\substack{+0.95 \\ -0.81}$ is in agreement with the SM expectation.
Despite the fact that much less data have been is used in this search, the results of the individual optimized and improved analyses, as well as their combination, achieved similar sensitivity and they are comparable to the Run I results.

\clearpage

\Acknowledgements
The acknowledgments go to the Swiss National Science Foundation, Switzerland and the Ministry of Education, Science and Technological Development, Serbia.

\end{document}